\documentclass[aps, prb, twocolumn, a4paper, superscriptaddress]{revtex4-2}

\usepackage{graphicx} 
\usepackage{dcolumn} 
\usepackage{amsmath, bm} 
\usepackage{hyperref} 
\usepackage{mathrsfs} 
\usepackage{color} 
\usepackage{gensymb} 
\usepackage{braket}
\usepackage[utf8]{inputenc}
\usepackage{tikz}
\usepackage{lipsum}
\usepackage{makecell}
\usepackage{multirow}

\makeatletter
\g@addto@macro\bfseries{\boldmath}
\makeatother

\def\mns{$M_{1/3}$NbS$_2$}
\def\mnsall{$M_{1/3}$NbS$_2$ ($M$ = Fe, V, Mn)}
\def\vns{V$_{1/3}$N\lowercase{b}S$_2$}
\def\mnns{M\lowercase{n}$_{1/3}$N\lowercase{b}S$_2$}
\def\fens{F\lowercase{e}$_{1/3}$N\lowercase{b}S$_2$}
\def\crns{C\lowercase{r}$_{1/3}$N\lowercase{b}S$_2$}
\def\crts{C\lowercase{r}$_{1/3}$T\lowercase{a}S$_2$}
\def\cotse{C\lowercase{o}$_{1/4}$T\lowercase{a}S\lowercase{e}$_2$}

\def\musr{$\mu$SR}

\begin{document}

\title{Magnetism in~\mnsall: insight into intercalated transition-metal dichalcogenides using~\musr}

\author{N.~P.~Bentley}
\affiliation{Department of Physics, Centre for Materials Physics, Durham University, Durham, DH1 3LE, United Kingdom}
\author{T.~L.~Breeze}
\affiliation{Department of Physics, Centre for Materials Physics, Durham University, Durham, DH1 3LE, United Kingdom}
\author{A.~Hern{\'a}ndez-Meli{\'a}n}
\affiliation{Department of Physics, Centre for Materials Physics, Durham University, Durham, DH1 3LE, United Kingdom}
\author{T.~J.~Hicken}
\affiliation{PSI Center for Neutron and Muon Sciences CNM, 5232 Villigen PSI, Switzerland}
\author{B.~M.~Huddart}
\affiliation{Oxford University Department of Physics, Clarendon Laboratory, Parks Road, Oxford OX1 3PU, United Kingdom}
\author{F.~L.~Pratt}
\affiliation{ISIS Neutron and Muon Source, STFC Rutherford Appleton Laboratory, Harwell, Didcot, OX11 OQX, United Kingdom.}
\author{A.~E.~Hall}
\affiliation{Department of Physics, University of Warwick, Coventry, CV4 7AL, United Kingdom.}
\author{D.~A.~Mayoh}
\affiliation{Department of Physics, University of Warwick, Coventry, CV4 7AL, United Kingdom.}
\author{G.~Balakrishnan}
\affiliation{Department of Physics, University of Warwick, Coventry, CV4 7AL, United Kingdom.}
\author{S.~J.~Clark}
\affiliation{Department of Physics, Centre for Materials Physics, Durham University, Durham, DH1 3LE, United Kingdom}
\author{T.~Lancaster}
\affiliation{Department of Physics, Centre for Materials Physics, Durham University, Durham, DH1 3LE, United Kingdom}

\begin{abstract}
We present the results of muon-spin relaxation (\musr) measurements of the static and dynamic magnetism of~\mnsall, three intercalated transition-metal dichalcogenides.
Transitions to long-range magnetic order are observed in all three materials and local magnetic fields at muon sites are compared to dipole field calculations.
Measurements on~\fens~capture the evolution of two coexisting  magnetic phases. 
In~\vns~we observe a peak in the dynamic response at $9$~K, coincident with previous reports of a possible low-temperature phase transition.
The observation of high-frequency muon precession in~\mnns~suggests the existence of an additional muon site that implies a difference in electronic energy landscape compared to the other materials in the series.
Taken together, this demonstrates that the change in intercalant species drives significant variations in magnetism, highlighting the~\mns~series as an ideal group of materials for investigating a wide range of magnetic phenomena.
\end{abstract}
\maketitle
\section{Introduction}
The transition-metal dichalcogenides (TMDCs) are a family of low-dimensional materials that exhibit a range of exotic electronic and optical properties~\cite{wilson1969transition,moncton1975study,rajora1987preparation,naito1982electrical,tissen2013pressure,witteveen2021polytypism,ramasubramaniam2011tunable,wang2012electronics}.
They are formed of hexagonal $AB_2$ layers, containing a transition metal, $A$, and a chalcogen, $B$. 
Due to the weak van der Waals bonding between layers, the resulting structures allow for a high degree of tunability, highlighted by the dependence of transport properties on sample thickness~\cite{castellanos2012laser,ganatra2014few}.   
The  properties of these materials can be further enhanced through the intercalation of transition metal ions, $M$, between the layers~\cite{parkin19803ii,parkin19803d,zhang2020chemrev,battaglia2007non,pan2023int,mayoh2022Cr1_3,parkin19803i,parkin1983magnetic,xie2022structure}. 
Specifically, a host of materials demonstrating a range of magnetic phenomena
are realized by varying the species and stoichiometry of the intercalate ~\cite{moriya1982evidence,edwards2023V1_3,little2020three,park2023Co1_3,dong2024Co1_3,han2017Cr1_3tri,yeochan2023Ni1_3,wu2023Fecharge,hawkhead2023intTMDC,mandujano2025Co_x}.
Here we focus on materials derived from NbS$_2$, a semiconductor with a temperature dependent bandgap~\cite{bharucha2022NbS2elec} that exhibits superconductivity below $5.7$~K in bulk~\cite{heil2017origin}, which is intercalated with concentration~$x=1/3$. 
Intercalated materials of the form~\mns~belong to the chiral space group $P6_322$ and have an electronic structure that can be understood via a simple band-filling mechanism \cite{hawkhead2023intTMDC}. 
However, their magnetic properties are diverse, showing a rich range of magnetic structures and excitations.

Our investigation centres on the system intercalated with Fe, V and Mn ions.
The model of band-filling used to describe the electronic structure of~\mns~also suggests that the magnetic exchange interaction changes with the filling of the intercalate 3$d$ band, with a less-than half-filled band leading to predominantly ferromagnetic (FM) interactions and a more-than half-filled band leading to predominantly antiferromagnetic (AFM) interactions.
\mnsall~are all close to this crossover, with the 3$d$ band slightly-less-than half-filled for the V material and slightly-more-than half-filled for the Fe material, so they will display particularly interesting and contrasting magnetic behaviours.
\fens~is characterized by a coexistence of two magnetic phases below $40$~K with distinct magnetic structures, that shows an evolution with temperature~\cite{wu2022Feconc}.~\vns~orders at $52$~K, and also shows evidence for a subtle change in behaviour around 10~K. Recently this has been linked to the emergence of non-Fermi liquid behaviour in the resistivity below this temperature, and it has been suggested that this reflects the existence of topological magnetic excitations in this regime~\cite{hall2021magnetic,ray2025_V}. 
Finally,~\mnns~displays ferromagnetic-like behaviour on the hundred nanometre length scale below $44$~K. This is possibly due to the presence of large-pitch-length chiral helimagnetism, as is the case in~\crns~\cite{miyadai1983magnetic,braam2015magnetic}.

Here we combine muon spin relaxation (\musr) measurements with density functional theory (DFT) calculations to  examine the changes in magnetism.
The~\musr~technique provides a powerful experimental tool for investigating magnetism in materials~\cite{blundell1999spin,hillier2022muon}. 
In particular, the local nature of the muon probe and the timescale of the dynamical response~\cite{muontextbook2022} provide insight into magnetic phenomena not easily accessible to other techniques.
Here we utilise this to probe the magnetism of~\mnsall, complemented by DFT calculations to determine candidate muon stopping sites for all three materials. 
\musr~has proven to be an effective tool for investigating the magnetism in intercalated TMDCs, with measurements on~\crns~and~\crts~identifying the presence of a chiral soliton lattice (CSL) and observing magnetic dynamics on the muon timescale consistent with a proposed pseudogap feature in the electronic structure~\cite{hicken2022energy}.
Muons have also been used to investigate~\cotse~\cite{graham2025_altermagnetism}, providing evidence for a proposed altermagnetic state and ruling out slow magnetic fluctuations as a contributing factor to the persistence of band spin-splitting into the paramagnetic phase.

This paper is structured as follows: 
In Section~\ref{ExpCom_detail} we describe the experimental and computational techniques underpinning our results; in Sections~\ref{Fe_sec}-\ref{Mn_sec} we analyse the results of~\musr~measurements on $M=$~Fe, V and Mn in turn, along with supporting DFT calculations of the muon stopping sites, before presenting our conclusions in Section~\ref{Conclusion_sec}.

\section{Experimental and computational details}
\label{ExpCom_detail}

\begin{figure}
  \centering
  \includegraphics[width=\linewidth]{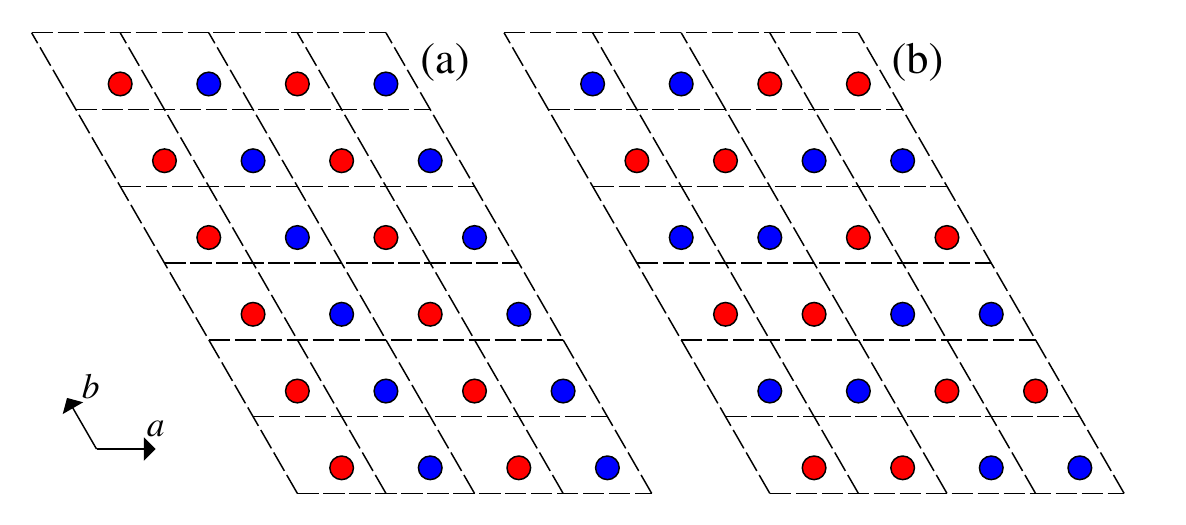}
  \caption{Magnetic structure of~\fens~for a slice in the $ab$ plane, with up spins (red) and down spins (blue). (a) Straight lines of a given spin orientation in the $b$ direction giving stripe ordering and (b) undulating lines in the $b$ direction giving zig-zag ordering.
  The spin of the second Fe ion in the unit cell of~\fens~ is given by shifting the magnetic structure one ion along $a$, producing alternating layers along the $c$ axis.}
  \label{fig:Fe_NbS2_magstr}
\end{figure}

\begin{figure*}
  \centering
  \includegraphics[width=1.0\linewidth]{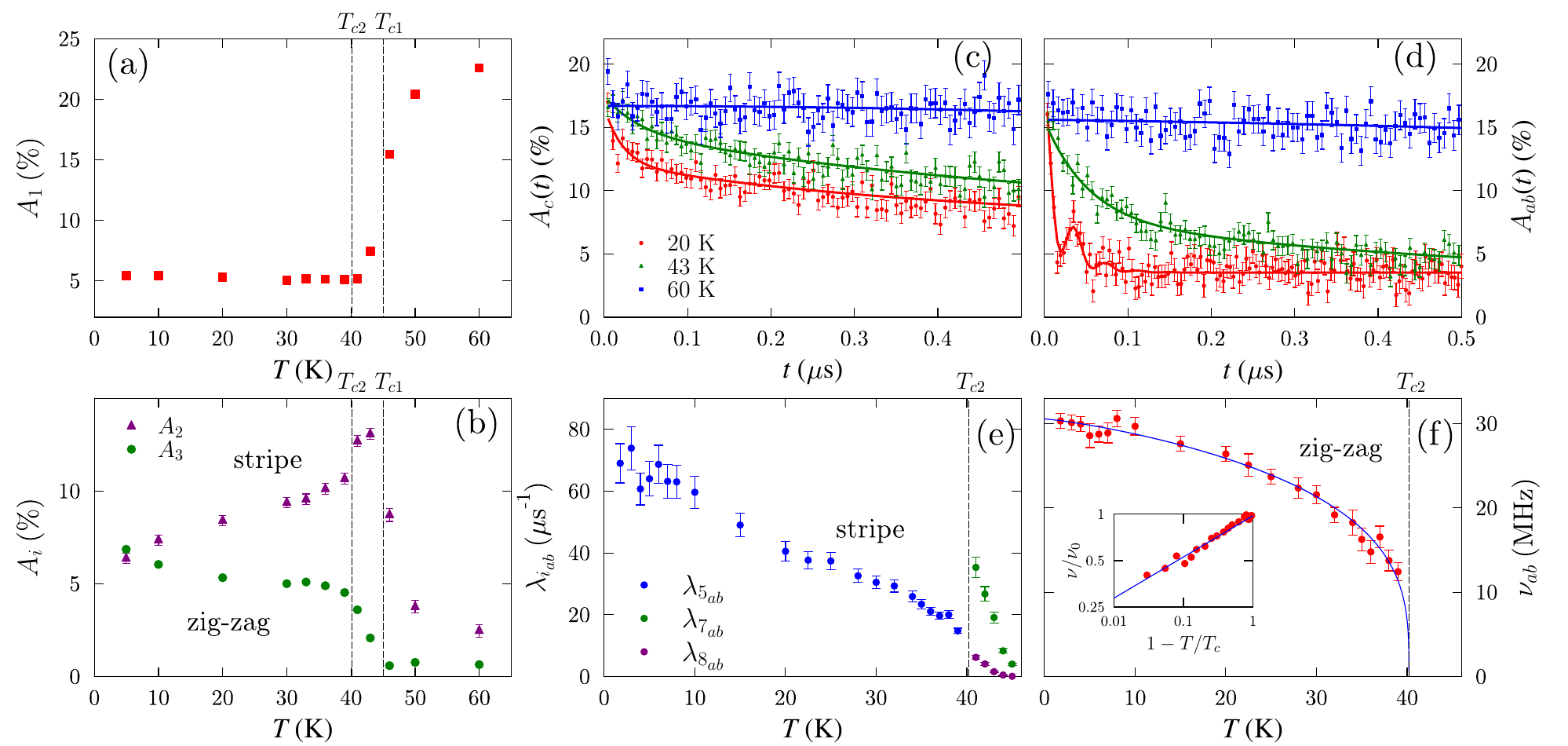}
  \caption{
  Temperature evolution of the wTF amplitudes associated with (a) non-magnetic regions of~\fens~and (b) the stripe, $A_2$ (purple triangles), and zigzag, $A_3$ (green circles), phases.
  Lines show the transition between the mixed and stripe-dominated regions at $40.2$~K and the transition to between stripe-dominated ordering and paramagnetism at $45$~K.
  ZF spectra for~\fens~measured at 20~K (red circles, mixed), 43~K (green triangles, stripe-dominated) and 60~K (blue squares, paramagnetic) for polarization along  (c) the $c$ axis and (d) the $a$-$b$ plane.
  (e) Temperature dependence of the relaxation rates $\lambda_{5_{ab}}$ (blue), $\lambda_{7_{ab}}$ (green) and $\lambda_{8_{ab}}$ (purple).
  (f) Temperature dependence of the precession frequency in the $a$-$b$ plane, with the fit described in the text.
  }
  \label{fig:FeNbS2_mu}
\end{figure*}
Single crystals of~\mnsall~were produced by the chemical vapour transport technique~\cite{edwards2024_trends,hall2022comparative}. 
Presynthesized polycrystalline powders of the correct composition were placed in an evacuated and sealed quartz tube with the transport agent (iodine).
One end of the tube was then heated to a maximum temperature of $950^\circ$C, with a difference of $100^\circ$C ($M=$~Fe and V) or $150^\circ$C ($M=$~Mn) across the tube.
After three weeks of heating the tubes were then cooled to room temperature, forming platelets ranging from $1$ to $5$~mm along their longest edges.
A backscattering X-ray Photonic-Science Laue camera system was used to assess the quality of the single crystals, while energy-dispersive X-ray spectroscopy analysis in a scanning electron microscope was used to estimate their composition: both techniques confirm the non-centrosymmetric $P6_322$ space group of the crystals.
The stoichiometry of the crystals has been confirmed through magnetic susceptibility measurements, as the critical temperature is sensitive to deviations from the expected intercalation concentration in the materials.

\musr~measurements on partially orientated mosaics of these single crystals were carried out using the GPS instrument~\cite{amato2017_GPS} at the Swiss Muon Source (S$\mu$S), Paul Scherrer Institute (PSI). 
The mosaics were aligned in an Ag foil envelope (foil thickness 25~$\mu$m) such that the $c$ axis of the crystals was collinear with the beam direction, before being mounted on a fork and inserted into a $^{4}$He cryostat. Depending on the initial muon polarization direction, the detector arrangement allows us separate access to the evolution of the muon spin polarization longitudinal to the beam direction (i.e.\ along the  crystallographic $c$ axis) and transverse (i.e.\ in the $a$-$b$ plane).
Additional measurements on a polycrystalline sample of the $N=$~V material were made at the STFC-ISIS facility, using the EMU spectrometer~\cite{giblin2014_EMU}.
Analysis of the muon data was carried out using the WiMDA program~\cite{pratt2000wimda}.

In a muon-spin relaxation experiment, spin-polarized positive muons are implanted in a sample and interact with the local magnetic fields. 
The angular distribution of the emitted decay positrons allows us to infer the spin polarization of the muon ensemble, via the measured positron asymmetry, $A(t)$.
For the measurements on~\vns~the muon spins were orientated along the $c$ axis of the crystal.
In contrast, for the measurements on~\fens~and~\mnns~the spins lie at a $45^\circ$ angle to the $c$ axis, so that the distinct components of the muon spin can be projected out, allowing for the observation of an anisotropic response  in the sample.

To determine muon stopping sites, we performed DFT geometry optimizations, using the MuFinder program~\cite{huddart2022mufinder}, based on the~\textsc{Castep}~\cite{clark2005first} plane-wave pseudopotential code.
We use the PBE functional and a supercell comprised of a muon (modelled by an ultrasoft hydrogen pseudopotential) placed in a $2\times2\times1$~cell of~\mns. 
Muonated supercells were generated by constraining the initial muon sites to be at least 1.0~\AA~away from atoms and at least 0.5~\AA~away from muons in previous supercells. 
The procedure was repeated for 30 initial muon locations in each material. 
Plane wave cutoffs of $800$, $900$ and $1000$~eV were used for the Fe, Mn and V intercalated materials. A $2\times2\times2$ Monkhurst-Pack grid for BZ sampling gave total energies that converge to within 10~meV per atom. Calculations were performed using lattice parameters consistent with experimentally measured values. 
The dipole field at each of the sites is calculated using  the expression
\begin{equation}
\label{eq:dipole_field}
    \textbf{B}_{\rm dip}(\textbf{r}_i) = \frac{\mu_0}{4\pi}\sum_{i} \left [\frac{(\textbf{m}_i\cdot\textbf{r}_i)\textbf{r}_i}{r_i^5} - \frac{\textbf{m}_i}{r_i^3}
    \right],
\end{equation}
for each possible magnetic configurations of the system, using the MuESR program~\cite{bonfa2018introduction}.
We include a Lorentz field, $B_{\rm L}$, due the finite size of the region which the dipole moments are summed over and a demagnetization field, $B_{\rm dem}$, which accounts for the size and shape of the sample. 
The sample is a flat plate normal to the $c$ axis (giving a demagnetisation tensor of $N_{cc}=1$ and $N_{\alpha\beta}=0$ otherwise). 
When the magnetic unit cell is distinct from the structural unit cell a muon implanted at a given site will experience a distribution of simulated magnetic fields, $p(B)$, rather than just a single field. 
The local field distribution $p(B)$ can then be compared to the frequency spectrum measured using~\musr.  
Muon implantation distorts the local crystalline environment of the muon site, altering the local magnetic field.
For~\mnsall~these muon induced distortions cause a significant difference in the field distribution produced by the dipole moments, so are included in our analysis.
Hyperfine coupling was estimated assuming an
interaction of the muon and an s-electron given by
\begin{equation}
\label{eq:hyperfine_contact}
B_{\rm cont}(\textbf{r}_{i})=\frac{2}{3}\mu_{\rm 0}\mu_{\rm B}\rho_{s}(\textbf{r}_{i}), 
\end{equation}
where $\rho_s$ is the electron spin density at the muon site~\cite{onuorah2018hf}.

\section{\fens}
\label{Fe_sec}
The low-temperature magnetism of~\fens~is characterised by the formation of a nematic state~\cite{little2020three,nair2020electrical,weber2021origins}, leading to the coexistence of two competing AFM phases.
The first is a layered AFM stripe phase [shown in Fig.~\ref{fig:Fe_NbS2_magstr}(a)], characterised by a propagation vector $\rm \textbf{k}_1$ = (0.5, 0, 0), with straight lines of a given spin orientation along $b$. The second is a layered AFM zig-zag phase with $\rm \textbf{k}_2$ = (0.25, 0.5, 0), in which there are undulating lines of a given spin orientation that alternate along $b$ and have an up-up-down-down arrangement along $a$ [Fig.~\ref{fig:Fe_NbS2_magstr}(b)]. 
Investigation of off-stoichiometry samples show that the AFM stripe is found for $x<1/3$, and the AFM zig-zag phase with $x>1/3$.
Neutron diffraction indicates that the magnetic moments in both structures are orientated along the $c$ axis, in agreement with the anisotropic magnetic response seen in bulk magnetometry~\cite{wu2022Feconc}.
Magnetometry and heat capacity measurements identify a pair of transitions at $T_{c2}=41$~K and $T_{c1}=45$~K, where the region between these two transitions hosts only the stripe phase, while below $T_{c2}$ both magnetic phases coexist.
The magnetic moments obtained from neutron scattering are $\mu_{\rm Fe}=2.9~\mu_{\rm B}$ for the stripe-dominated region and $\mu_{\rm Fe}=3.3~\mu_{\rm B}$ for the mixed region.

Analysis of the neutron measurements suggested a non-trivial evolution of the coexisting magnetic phases with temperature. 
In order to track the change in volume fraction of each of these phases, we performed weak transverse-field (wTF) $\mu$SR measurements on~\fens~in an applied field $B_{\rm wTF}=5$~mT.
The resulting asymmetry is fitted with
\begin{equation}
\label{eq:Fe_wTF}
A(t) = A_1\cos(\gamma_\mu B_{\rm wTF} t)e^{-\lambda_1t} + A_2e^{-\lambda_2t} + A_{3}e^{-\lambda_3t},
\end{equation}
where $\lambda_1=0.11(1)$~$\rm\mu s^{-1}$, $\lambda_2 = 2.62(7)$~$\rm\mu s^{-1}$ and $\frac{\gamma_\mu}{2\pi}B_{\rm wTF}=0.673(1)$~MHz are globally refined.
The component corresponding to muons implanting in non-magnetic environments, $A_1$, is constant at low temperatures and increases to a maximum above the magnetic ordering temperature $T_{c1}$.
The change in amplitude occurs over a wide temperature range above the transition [Fig.~\ref{fig:FeNbS2_mu}(a)], suggesting the persistence of quasistatic magnetic correlations above $T_{c1}$.

The amplitudes $A_2$ and $A_3$ reflect those muons stopping in regions of magnetic order. Both amplitudes decrease as the temperature is increased above $T_{c1}$.
There is a peak in $A_2$ in the stripe-dominated region (followed by a rapid decrease above $43$~K), and $A_3$ has a possible inflection around $30$~K [Fig.~\ref{fig:FeNbS2_mu}(b)].
These amplitudes indicate the proportion of each magnetic volume fraction present in the sample. Comparing these with neutron peak intensities~\cite{wu2022Feconc}, we note that the trend in behaviour of $A_2$ closely matches the neutron intensity attributed to the stripe phase, while $A_3$ follows that of the zig-zag phase.
The temperature dependence then suggests that the volume fraction of each coexisting phase is approximately equal at 5~K, but evolves roughly linearly such that at $T_{c2}$ the ratio of stripe phase to zig-zag phase is roughly 2:1. In the region $T_{c2}<T<T_{c1}$ the volume of zig-zag phase drops rapidly to zero on warming, although it is notable that although the stripe phase is the majority phase, a small contribution from the zig-zag phase persists.
The changes in $A_2$ and $A_3$ are consistent with the phase transition temperatures seen in magnetometry and in our ZF~\musr~measurements (see below). However, it is notable that they occur around $10$~K higher in temperature than the similar behaviour in the peak neutron scattering intensities. This possibly reflects the distinct length scales probed by the different techniques, with the muon results suggesting that regions of relatively short-range order (on a scale of order ten lattice spacings) persist up to the observed transitions. 

When both magnetic phases coexist we observe an anisotropy in the muon response
between the muon spin polarization projected along $c$ and that in the $a$-$b$ plane. 
The ZF asymmetry in the $c$-axis direction is described by the purely relaxing function
\begin{equation}
A_c(t) = A_{4_c}e^{-\lambda_{4_c}t} + A_{5_c}e^{-\lambda_{5_c}t} + A_{6_c},\label{eq:c}
\end{equation}
where $A_{4_c}=4.8(1)\%$, $A_{5_c}=4.4(2)\%$, $A_{6_c}=7.4(1)\%$ and $\lambda_{4_c}=2.4(2)~\rm\mu s^{-1}$ are globally refined.
The ZF asymmetry in the $a$-$b$ plane is described by the oscillatory function
\begin{equation}
A_{ab}(t) = A_{4_{ab}}\cos(2\pi\nu t)e^{-\lambda_{4_{ab}}t} + A_{5_{ab}}e^{-\lambda_{5_{ab}}t} + A_{6_{ab}},
\label{eq:ab}
\end{equation} 
where $A_{4_{ab}}=6.0(3)\%$, $A_{5_{ab}}=9.6(2)\%$, and $\lambda_{4_{ab}}=39(2)~\rm\mu s^{-1}$ are globally refined.
In both Eq.~\ref{eq:c} and Eq.~\ref{eq:ab}
the final terms describe a background signal from muons implanting in non-magnetic environments.
For the $c$-axis polarization, the size of the relaxation rate in the first term in Eq.~\ref{eq:c} is typical of the relaxation of muons due to slow magnetic dynamics. 
For the $a$-$b$ polarization in Eq.~\ref{eq:ab}, the first term describes muons precessing in a local field at a frequency $\nu$.

Using a scaling function for the fitted frequency of $\nu = \nu(0)\left[1-\left(T/T_{c2}\right)^\alpha\right]^\beta$, we obtain a transition temperature of $T_{c2} = 40.2(4)$~K, parameter $\alpha=1.2(3)$ and critical exponent $\beta=0.27(2)$.
This critical temperature $T_{c2}$ is consistent with the reported transition to stripe-dominated magnetism~\cite{wu2022Feconc} and the onset of the decrease of $A_3$ in our wTF measurements.
The parameter $\beta$ is closer to that of the 3D Ising model ($0.32$) than 3D Heisenberg model ($0.367$), which is reasonable given the Ising-like alignment of spins along the $c$ axis in the zig-zag structure. 

Having linked $A_4{_{ab}}$ to the precession of muons in the zig-zag phase, the absence of an oscillating signal in the $c$-axis polarization 
suggests that the slowly relaxing term with amplitude $A_{4_c}$  
arises from
muon-spin components parallel to the local field in regions of zig-zag order.
Support for this comes from the relaxation rates, $\lambda_{4_c}$ and $\lambda_{4_{ab}}$, 
which differ by an order of magnitude, 
with the larger rate
$\lambda_{4_{ab}}$ corresponding to 
relaxation due to both fluctuations in the local magnetic order and magnetic dynamics.

The $A_{5_{\alpha=c,ab}}$ components account for muon spins that that are relaxed in regions of stripe magnetic order. For this phase,  oscillations due to LRO are not observed, suggesting that
magnetic disorder in the stripe phase washes out coherent precession of the muon polarization.
Temperature variation in the  the local field distribution 
is captured by the relaxation rate $\lambda_{5_\alpha}$ [see Fig.~\ref{fig:FeNbS2_mu}(e)].
Deviations from order-parameter-like behaviour (such as the reduced rate of change of $\lambda_{5_{ab}}$ between $20$ and $30$~K) can be attributed to changes in the nature of the field distribution and the magnetic dynamics with temperature.

When the stripe phase is dominant above $T=40$~K, the anisotropy in muon relaxation vanishes and the ZF asymmetry in both directions is described by
\begin{equation}
A_\alpha(t) = A_{7_\alpha}e^{-\lambda_{7_\alpha}t} + A_{8_\alpha}e^{-\lambda_{8_\alpha}t} + A_{9_\alpha},
\end{equation}
where $\alpha=ab,~c$ and $A_{7_c}=2.9(3)\%$, $A_{7_{ab}}=7.9(5)\%$, $A_{8_c}=7.3(3)\%$, $A_{8_{ab}}=5.4(4)\%$, $A_{9_c}=7.34(3)\%$ and $A_{9_{ab}}=2.2(1)\%$ are globally refined.
The relaxation is captured  by a fast ($\lambda_7$) and a slow ($\lambda_8$) contribution, both of which now decrease with temperature.
This behaviour can partly be attributed to a decrease in the local field strength at the muon site, although it also reflects the change in volume fraction of the two phases, which is not accounted for in this fitting routine. 
The fitting (for both the stripe-dominated and the mixed region) was also carried out by fixing the amplitudes based on the volume fractions measured in the wTF measurements.
This resulted in similar behaviour in the relaxation rates and the precessions frequency, so we can be confident fitting with global refined amplitudes captures the variations in magnetic relaxation in~\fens.
It is notable that the slow relaxation from the stripe phase $\lambda_{7}$ continues the trend seen in $\lambda_{5}$ below $T_{c2}$. The emergence of a large relaxation rate $\lambda_{7}$ implies that the muons in sites in the former zig-zag regions are strongly depolarized, presumably by disorder of the Fe spins.

Finally, the spectra above the transition temperature for both polarization directions are described by 
\begin{equation}
\label{eq:N_ZFhighT}
A(t) = A_{10_\alpha}e^{-\lambda_\alpha t} + A_{11_\alpha}e^{-\sigma_{\alpha}^2t^2} + A_{12_\alpha},
\end{equation}
where $\lambda,~\sigma$ and $A_{12_\alpha}$ are held constant.
The exponential term represents the relaxation of muon spins due to magnetic  dynamics and the Gaussian term represents the relaxation due to quasistatic disordered magnetic moments.
We also note here that the spectra measured at temperatures above the magnetic transition for~\vns~and~\mnns~are described by the same functional form.

Muon site calculations allow us to identify a cluster of candidate low-energy sites at the centre of a triangle of Nb atoms, in agreement with sites identified in other members of the series~\cite{hicken2022energy}, that we call A, B and C. 
They are crystallographically equivalent in the undistorted crystal (through the action of the symmetry operators of the $P6_322$ space group, Table~\ref{tab:sym_muonsite}), but can be magnetically inequivalent.
In~\fens~the candidate site realized at C is at least 0.3~eV lower in energy than any of the other sites, due to the local spin-density supporting muon induced distortions that form a lower energy arrangement of atoms, breaking local site symmetry.
\begin{table}
\begin{tabular}{c|c|c}
       Site & Coordinates & Wyckoff Site~~\\
    
       \hline
         A & (2/3, 2/3, 0) & $6g$~\\
         B & (1/3, 0, 0) & $6g$~\\
         C & (0, 1/3, 0) & $6g$~\\
         D & (3/4, 2/3, 1/4) & $12i$~\\
         E & (11/12, 1/4, 1/4) & $12i$~\\
\end{tabular}
\caption{Location and symmetry of candidate muon sites calculated for~\mns.}
\label{tab:sym_muonsite}
\end{table}
\begin{figure}
  \centering
  \includegraphics[width=0.7\linewidth]{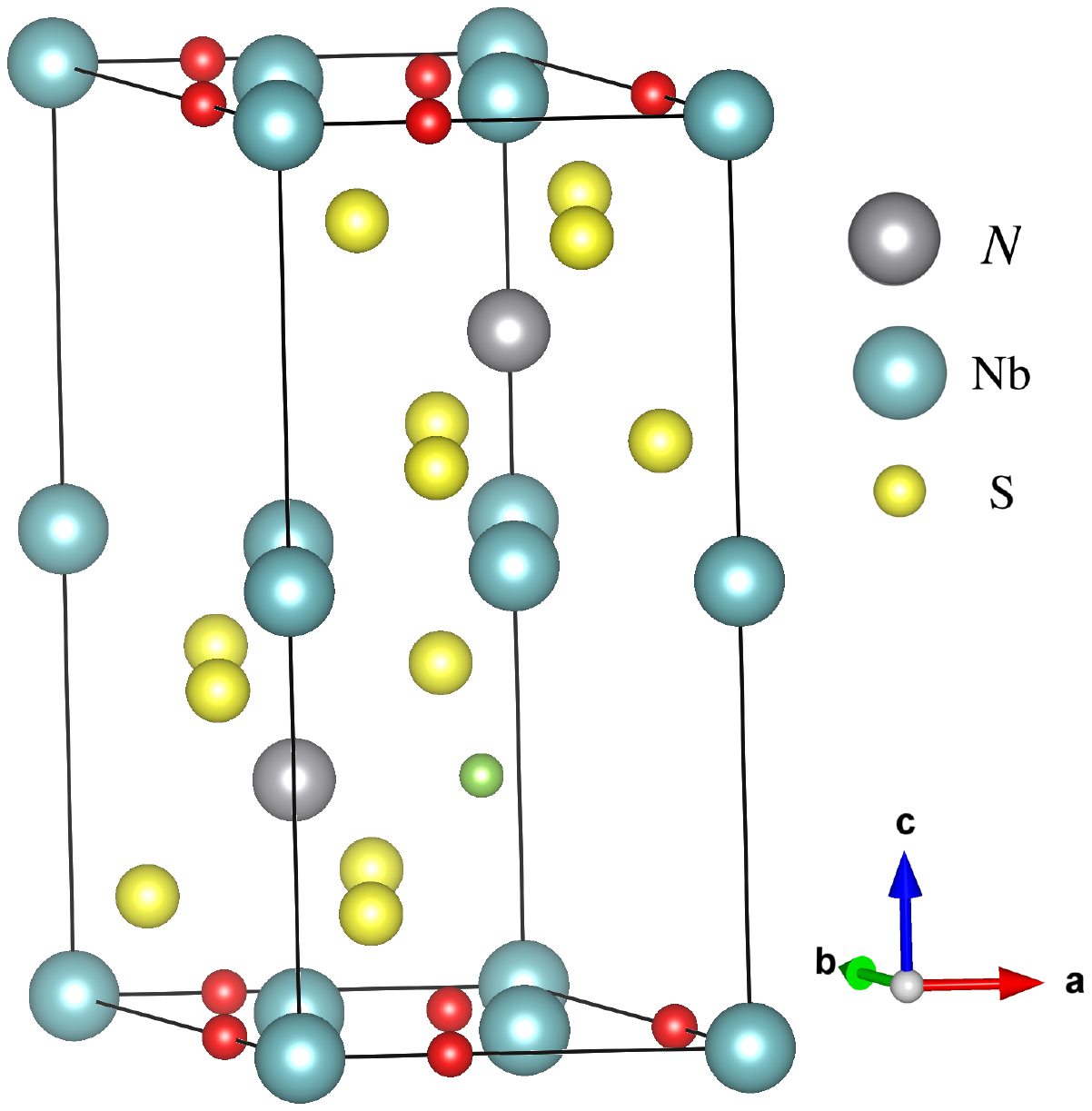}
  \caption{Location of candidate muon sites in~\mnsall, with the three low energy sites at the centre of triangles of 3 Nb ions shown in red and the high-energy muon site in~\mnns~shown in green.}
  \label{fig:MNbS2_muonsite}
\end{figure}

Dipole field calculations for the AFM stripe structure in the $M=$~Fe material, assuming muons implanted at site C, result in a single peak in the simulated magnetic field distribution $p(B)$ at $\approx 36$~MHz in the mixed region, and $\approx32$~MHz in the stripe-dominated case. The AFM zig-zag ordering produces a cluster of peaks tightly centred around $20$~MHz in $p(B)$.
The behaviour of our wTF amplitudes suggests phase separation between regions of stripe and zig-zag magnetic ordering, so our dipole fields have been calculated for each magnetic structure assuming the local vicinity of the implanted muon belongs to the same phase.
Phase separation would be expected to produce at least two precession frequencies in the~\musr~spectra when both phases coexist, so the observation of a single precession frequency supports the hypothesis that oscillations occur due to muon precession in only one of the phases.  
To obtain agreement with our measurements, assuming the oscillations occur due to muons implanted in regions of zig-zag ordering, we have to increase the moment measured with neutrons by a scale factor of 1.5.
This implies a value of $4.9~\mu_{\rm B}$, which is in agreement with value of the spin-only moment for $S=2$ Fe$^{2+}$ ions and susceptibility measurements~\cite{haley2020half}.
Moreover, this difference isn't accounted for by the calculated hyperfine contact field at this site [see Tab.~\ref{tab:hyperfine}].
For the stripe structure the calculated fields are orientated along $[0.7, 0.4, 0.55]$, which agrees with the observation of similar muon relaxation along the $c$ axis and in the $a$-$b$ plane.
Meanwhile, for the zig-zag structure the fields are orientated almost along $[0,0,1]$, consistent with the observation of relaxation due to magnetic dynamics in the $c$ direction and precession of the muon in the local field in the $a$-$b$ plane.

To summarize, we have used \musr~to confirm the existence of low-$T$ phase separation, where stripe and zig-zag magnetic order coexist, along with an intermediate temperature region dominated by the stripe phase. 
We attribute oscillations in our measured spectra to muons implanting in regions of zig-zag order. The associated anisotropy in the muon relaxation is then consistent with the orientation of the local fields at the muon site for both magnetic phases. 

\section{\vns}
\label{V_sec}
\vns~has attracted recent interest due to low-temperature behaviour suggesting the presence of topological magnetism~\cite{ray2025_V}.
Initially magnetometry measurements on a powder sample identified~\vns~to be paramagnetic~\cite{hulliger1970magnetic}, before measurements on single crystals suggested FM~\cite{parkin19803i} order within the $a$-$b$ plane.
Single crystal neutron diffraction identified weakly canted AFM~\cite{lu2020V1_3,ray2025_V}, with a transition to LRO at 50.1(1)~K.
Refinement of the neutron scattering data results in a simple AFM structure formed from FM layers of V$^{3+}$ moments, and any canting is constrained so the net moment along the $c$ axis is $<0.1\mu_{\rm B}$ per V ion. 
The moments are orientated along the $a$-axis with $\mu_{\rm V}=1.7(1)~\mu_{\rm B}$ and the resulting magnetic structure is described by a $\rm\textbf{k}_{0}$ = (0, 0, 0) propagation vector.
Other neutron scattering measurements on single crystal and powder samples of~\vns~suggests that the ordered magnetic state realizes a double-Q structure~\cite{hall2021magnetic}, where $\rm \textbf{k}_0$ = (0, 0, 0) corresponds to an AFM structure formed from FM layers of V$^{3+}$ moments orientated along the $a$-axis (determined via symmetry analysis) with $\mu_{\rm V_{\it ab}}=0.90(5)~\mu_{\rm B}$, and $\rm \textbf{k}_1$ = (0, 0,~$\frac{1}{3}$), describes a longitudinal spin density wave (SDW) along $c$~\cite{kimura2008longspinden}. 
This enforces an up-down-down pattern for the moments, where $\mu_{\rm V_{\it c}}$ equals $1.12(12)~\mu_{\rm B}$ and $0.61(6)~\mu_{\rm B}$ for the up and down moments respectively.
It is unclear whether the propagation vectors describe the same magnetic phase or two distinct, coexisting magnetic phases; diffuse magnetic scattering is also present below $T_c$. 

A low-temperature magnetic transition was suggested to occur around~$10$~K~from magnetometry and neutron diffraction measurements~\cite{parkin19803i,hall2021magnetic,ray2025_V}, although its properties remained mysterious.
Recently, this low-$T$ transition has also been  observed in a sizeable  spontaneous anomalous Hall effect (AHE), similar to that seen in topological magnetic materials. Moreover, this is accompanied by a region of low-temperature non-Fermi liquid (NFL) behaviour in the ZF resistivity, $\rho(T)$~\cite{ray2025_V}. The presence of both the spontaneous AHE and NFL behaviour suggests the AFM domain walls that form in~\vns~are topological magnetic excitations. 
Similar topological excitations have previously been observed in~\crns, where the application of small magnetic field forms a CSL~\cite{togawa2012chiral,hicken2022energy}.

\begin{figure}
  \centering
  \includegraphics[width=1\linewidth]{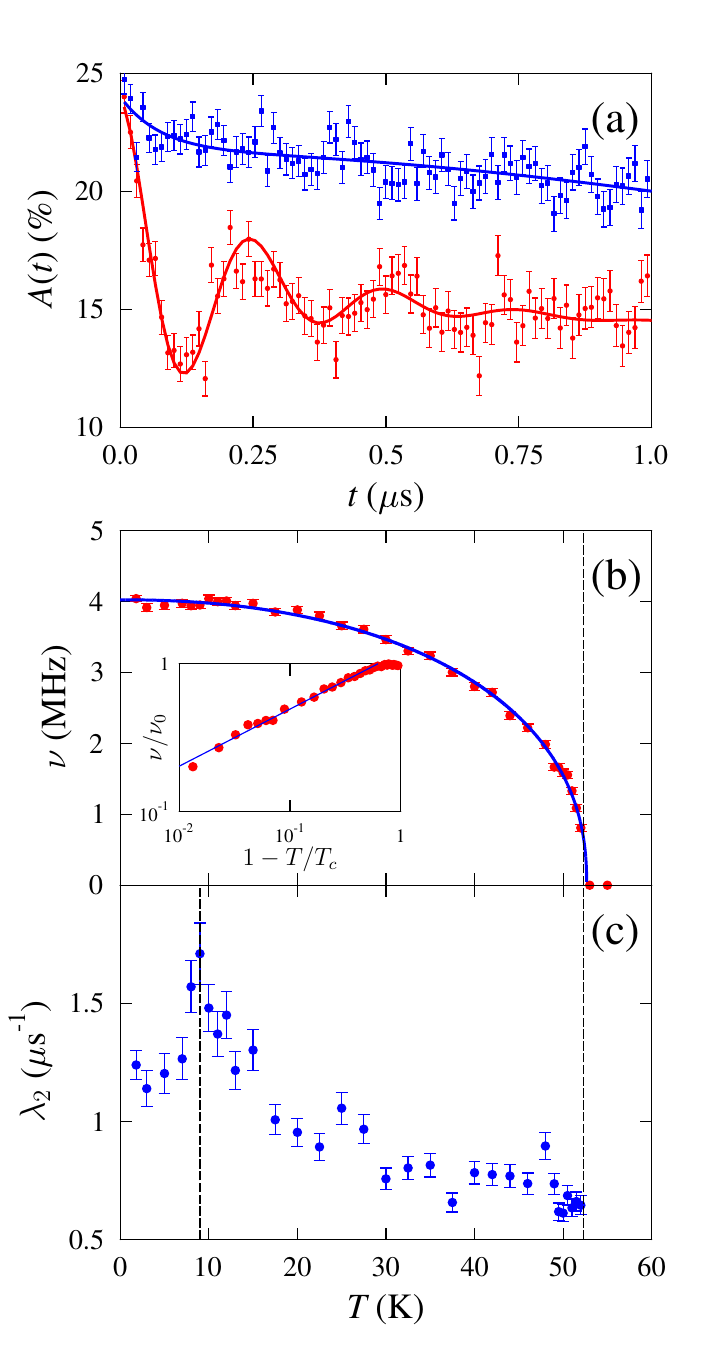}
  \caption{(a) ZF spectra for~\vns~measured at 5~K (red circles) and 80~K (blue squares).
  (b) Temperature dependence of the precession frequency, with a fit of critical scaling shown in blue. 
  (c) Dynamical relaxation rate of the $A_2$ component of the asymmetry, showing a peak at $9$~K. 
  }
  \label{fig:VNbS2_ZF}
\end{figure}

Example ZF $\mu$SR spectra are shown in Fig.~\ref{fig:VNbS2_ZF}. 
Those measured below the transition to LRO are captured by
\begin{equation}
A(t) = A_1\cos(2\pi\nu t)e^{-\lambda_1t} + A_2e^{-\lambda_2t} + A_{3},
\end{equation}
where $A_1=7.33\%$, $A_2=3.12\%$, $A_3=13.55\%$ and $\lambda_1=5~\rm\mu s^{-1}$ are held constant.
The first term accounts for the component of muon spin initially perpendicular to the local field at the muon implantation site, which precesses coherently at a frequency $\nu$. 
The second term describes the muon spin component initially parallel to the local field, so is relaxed only by dynamics, and the third term corresponds to muons stopping in non-magnetic environments. 
We find a critical temperature of $T_c = 52.7(2)$~K, parameter $\alpha=2.2(2)$ and critical exponent $\beta = 0.38(1)$ [see Fig.~\ref{fig:VNbS2_ZF}(b)], which is similar to the expected value for 3D Heisenberg fluctuations ($\beta = 0.367$).
We carried out wTF muon measurements that provide no evidence of phase separation, suggesting the double-Q structure would be realised in a single magnetic phase.

Sites A, B and C [Table~\ref{tab:sym_muonsite}] are found in our muon site calculations and form a cluster $0.7$~eV lower in energy than other candidate sites.
We can calculate the dipole fields at each of the sites for both the simple AFM and double-Q structures and compare to the amplitudes measured in ZF~\musr,
which can be expressed in terms of dipole field components using $A_1 = A_\perp \propto \sum_i \left(\frac{B_{a,i}^2+B_{b,i}^2}{B_i^2}\right)p_i$ and $A_2 = A_\parallel \propto \sum_i \left(\frac{B_{c,i}}{B_i}\right)^2p_i$, where $\parallel$ and $\perp$ 
describe
the orientation of the components of the muon spin relative to the local field direction and $p_{i}$ is the fraction of muon sites experiencing the field $B_{i}$~\cite{huddart2025_GdRu2Si2}. 
For both magnetic structures $A_\perp\propto0.7$ and $A_\parallel\propto0.3$, agreeing with the ratio of the experimentally measured amplitudes, $A_1=7.33\%$ and $A_2=3.12\%$.

\begin{figure}
  \centering
  \includegraphics[width=1.0\linewidth]{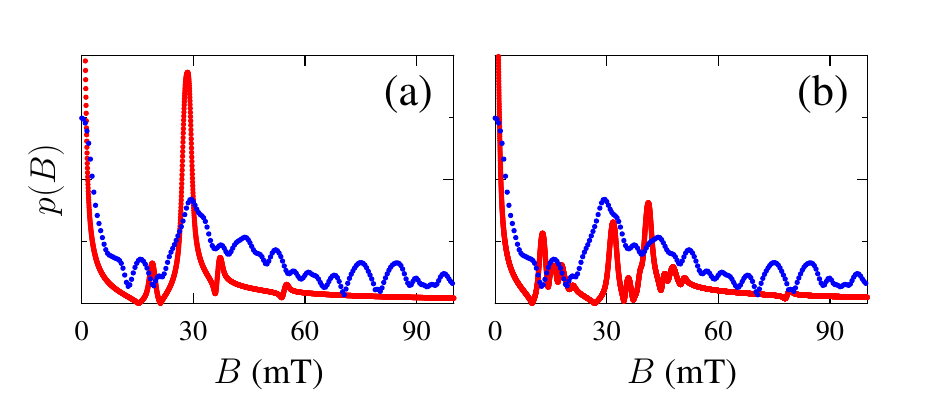}
  \caption{Simulated field spectra (red) for (a) the simple AFM and (b) the double-Q  magnetic structures in the $N=$V material, compared to the field distribution measured using~\musr~(blue). The moment from neutron measurements has been scaled by a factor of $1.5$ for simulations of the double-Q spectra and all the spectra are normalised.
  }
  \label{fig:VNbS2_fieldspectra}
\end{figure}

The simple AFM structure shows a peak in the computed probability function $p(B)$ at the field value $B$ seen in experiment, but the peak is narrower than in the experimental spectrum. 
Calculations for the double-Q structure using moment values from neutron scattering underestimate the magnitude of field values of the peaks in the distribution by a factor of $1.5$ (i.e.\ similar to the scaling required for the Fe material). 
Scaling with this factor gives a moment value of $2.25~\mu_{\rm B}$, which is still less than the spin-only moment for $S=1$ V$^{3+}$ ions ($2.83~\mu_{\rm B}$) and moment from susceptibility measurements ($2.90~\mu_{\rm B}$)~\cite{hall2021magnetic}.
Additionally, the hyperfine contact field for the candidate muon sites in~\vns~is approximately zero [Tab.~\ref{tab:hyperfine}], so cannot account for the discrepancy in the size of the calculated and measured local field.
However, the width of $p(B)$ for the double-Q structure more closely matches the single broad peak seen in the~\musr, due in part to the larger number of possible field values for each site [see Fig.~\ref{fig:VNbS2_fieldspectra}(b)].
Additionally $p(B)$ changes negligibly when the moments are canted along the $c$ axis within the experimentally determined limit ($0.1\mu_{\rm B}$ per V ion).  
This suggests that canting is not responsible for the low-$T$ transition, in agreement with conclusions from previous work based on neutron measurements~\cite{ray2025_V}. 
Our conclusion is that the single broad frequency that we observe is compatible with both proposed magnetic structures.

The behaviour of the magnetic dynamics in~\vns~is quantified by the dynamical relaxation rate $\lambda_2$ [Fig.~\ref{fig:VNbS2_ZF}(c)]. Our main result for this material is that this quantity peaks at $9$~K, suggesting a change in magnetic dynamics on the muon timescale at the low-$T$ transition.
AC susceptibility measurements sensitive to slower magnetic dynamics also capture this low-$T$ transition, with an increase in $\chi'$ below $10$~K~\cite{Hallthesis}. 
These features in the magnetic dynamics are coincident with the onset of NFL temperature scaling in $\rho(T)$, a possible mechanism for which is the formation of a chiral spin texture~\cite{ray2025_V}.
Such a texture consists of topological magnetic objects that exhibit magnetic dynamics to which the muons are sensitive.
In particular, the behaviour of $\lambda$ at the upper boundary of the NFL region is reminiscent of the increase of the relaxation rate seen due to skyrmion lattice formation in Cu$_2$OSeO$_3$~\cite{hicken2021_dynamics}.

\section{\mnns}
\label{Mn_sec}
\begin{figure}
  \centering
  \includegraphics[width=0.9\linewidth]{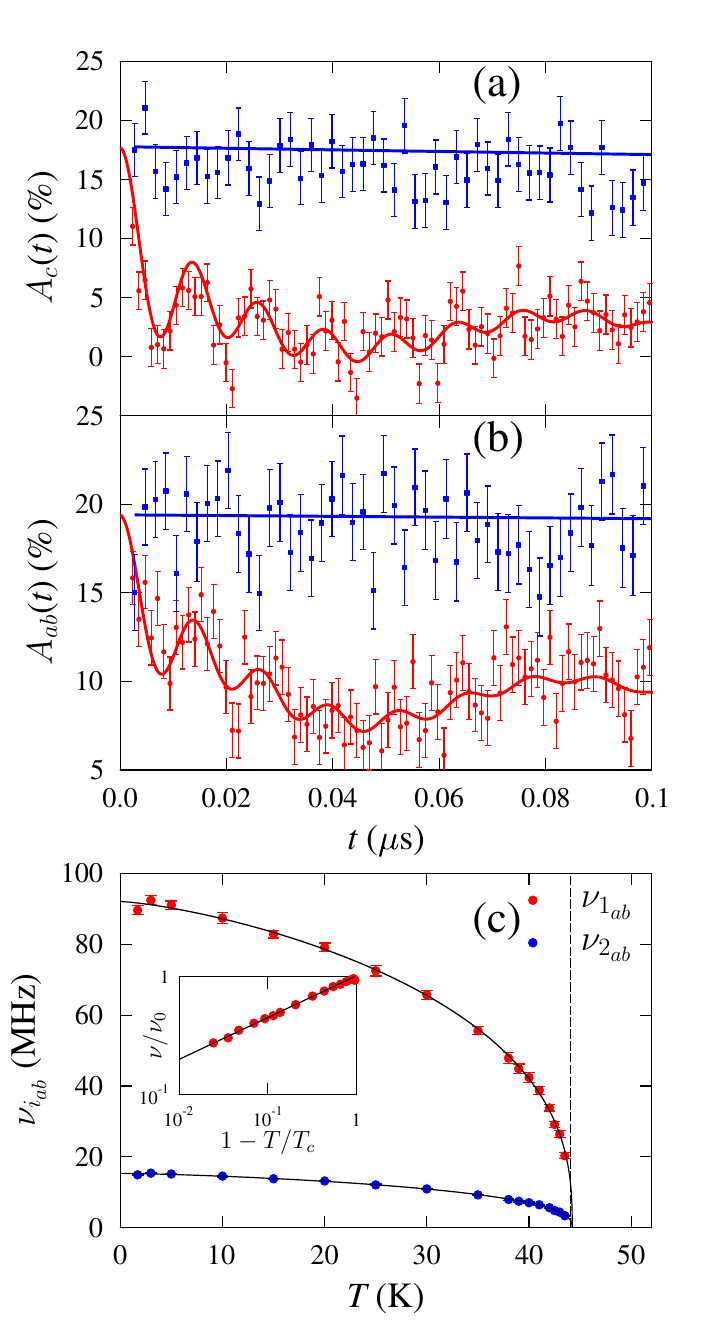}
  \caption{ZF spectra for~\mnns~corresponding to the detector pair along the $c$ axis (a) and in the $a$-$b$ plane (b) of the crystal, measured at 20~K (red circles) and 60~K (blue squares) which are below and above the transition to magnetic order respectively. 
  Temperature dependence of the precession frequencies measured for spectra in the $a$-$b$ plane (c), where the fit of the critical scaling function is shown in black.  
  }
  \label{fig:MnNbS2_ZF}
\end{figure}
The possible presence of helimagnetism (similar to~\crns~\cite{hall2022comparative,sahoo2024nmr}) and differences in the underlying electronic structure result in the magnetism of~\mnns~differing from the Fe and V intercalated materials. 
Studies involving magnetic susceptibility, neutron scattering and Lorentz transmission electron microscopy suggest the presence of ferromagnetic-like behaviour on the local length scale~\cite{karna2019Mn1_3chiral,dai2019Mn1_3,hall2022comparative},
which could be caused by ferromagnetic domains of size $\approx250$~nm along the $c$ axis.
These domains are separated by either topologically trivial domain walls ($\pi$-spin flips) or chiral soliton kinks ($2\pi$-spin flips), and the Mn moments are orientated along the [1 2 0] direction~\cite{hall2022comparative}.
Alternatively, the local ferromagnetism is also consistent with a helical magnetic structure propagating along the $c$ axis with a large pitch length ($\lambda\approx2500$~\AA) and Mn moments in the $a$-$b$ plane, which is expressed in propagation vector notation as $k = (0, 0, 0.005)$.
Recent NMR measurements~\cite{sahoo2024nmr} demonstrate anisotropy suggestive of a chiral helical majority phase, similar to the state observed in~\crns~\cite{hicken2022energy,togawa2012chiral}.
Single-crystal neutron scattering and DFT calculations~\cite{karna2019Mn1_3chiral,sahoo2024nmr}
give a moment size of~$\mu_{\rm Mn}=4.3(2)~\mu_{\rm B}$. 

Example $\mu$SR spectra are shown in Fig.~\ref{fig:MnNbS2_ZF} which, in contrast to other materials in the series, show two precession frequencies.
At temperatures below the transition to LRO, the asymmetry is found to be anisotropic and is described by the relaxation functions
\begin{equation}
A_c(t) = A_{1_c}J_0(2\pi\nu_{1_c} t)e^{-\lambda_{1_c}t} + A_{2_c}J_0(2\pi\nu_{2_c} t) + A_{3_c},
\label{eq:mn1}
\end{equation}
and 
\begin{align}
A_{ab}(t) &= A_{1_{ab}}J_0(2\pi\nu_{1_{ab}} t)e^{-\lambda_{1_{ab}}t} + A_{2_{ab}}J_0(2\pi\nu_{2_{ab}} t) \nonumber\\
&+ A_{3_{ab}}e^{-\lambda_{3_{ab}}t},
\label{eq:mn2}
\end{align}
where $A_{1_c}=11.4(3)\%$, $A_{1_{ab}}=6.3(3)\%$, $A_{2_c}=3.1(1)\%$, $A_{2_{ab}}=3.7(1)\%$, $A_{3_c}=2.32(5)\%$, $A_{3_{ab}}=9.41(3)\%$, $\lambda_{1_c}=11.4(2)~\rm\mu s^{-1}$, $\lambda_{1_{ab}}=14(3)~\rm\mu s^{-1}$ and $\lambda_{3_{ab}}=0.45(1)~\rm\mu s^{-1}$, are globally refined. 
The $A_{3_c}$ component is a constant background, corresponding to muons implanting in non-magnetic regions like the silver foil packet and the cryostat tails, while the $A_{3_{ab}}$ component is a slowly relaxing background that remains unchanged with temperature, corresponding to muons relaxed by slow magnetic dynamics.
The first two components in both Eqs.~\ref{eq:mn1}  and \ref{eq:mn2} correspond to oscillations described by zeroth-order Bessel functions of the first kind, $J_0$, capturing the muon response due to incommensurate ordering \cite{muontextbook2022}.
The frequencies of the observed oscillations can be fixed in proportion such that $\nu_1 = 6\nu_2$, suggesting a uniform decrease in the strength of the magnetic ordering across muon sites, consistent with a single magnetic phase. 
Fitting has also been performed using the helical model described in~\cite{amato2014_MnSi}. However, the dependence of this model on both the average field value and width of the underlying field distribution at every temperature (which is unknown) means we are unable to extract the observed order-parameter-like behaviour of the precession frequencies using this model.
wTF~\musr~measurements have also been performed on~\mnns~and they show no evidence of phase separation.

Treating the muon precession frequency as an order parameter 
gives a transition temperature of $T_{c}=44.2(3)$~K, parameter~$\alpha=1.4(2)$ and critical exponent $\beta=0.35(1)$.  
The value of the critical temperature $T_c$ is in agreement with previous susceptibility measurements~\cite{hall2022comparative,karna2019Mn1_3chiral} and the  exponent is similar to the expected value for 3D Heisenberg fluctuations ($\beta=0.367$).

In agreement with other materials in the~\mns~series, muon site calculations identify a cluster of low energy sites at the centre of three Nb atoms.
However, in contrast to~\fens~and~\vns, our DFT calculations are unable to reproduce candidate site A, (2/3, 2/3, 0), in~\mnns.
For the FM state, the calculated dipole field splits the low energy sites into two groups.
One of the groups has a local field that corresponds to a precession frequency of $\approx16$~MHz, while the second group of sites 
has a precession frequency of $\approx10$~MHz (although this latter group is only present when muon induced distortions are included).
A frequency of $\nu_2\approx16$~MHz is seen in our~\musr~measurements, matching the first group of sites. 
For the helical magnetic state implanted muons will experience a distribution of magnetic fields, $p(B)$ due to FM aligned moments in all possible orientations within the $a$-$b$ plane. 
This results in bimodal field distributions (regardless of whether muon induced distortions are included), with the peaks corresponding to precessions frequencies of $\approx4$~MHz and $\approx16$~MHz, the second of which matches $\nu_2$.
Simulations of $p(B)$ have used the magnetic moment measured by neutrons, $\mu_{\rm Mn}=4.3(2)~\mu_{\rm B}$, as it demonstrates the best agreement with $\nu_2$. 
In contrast, the spin-only moment of $4.90~\mu_{\rm B}$ for an ion valence state of Mn$^{3+}$ and magnetometry measurements of $\mu_{\rm Mn}=4.97~\mu_{\rm B}$ and $\mu_{\rm Mn}=5.43(1)~\mu_{\rm B}$ for $H\perp c$~\cite{karna2019Mn1_3chiral,hall2022comparative} all give a larger than observed field at the low energy muon site.

For the magnetic structures considered, it is impossible to realise the larger frequency, $\nu_1\approx 90$~MHz, at any of the low energy sites.
One possible explanation is that a higher-energy muon site is realised in~\mnns.
Higher-energy candidate muon sites in close proximity ($\approx1.4$~\AA) to a single sulfur ion [see the green sphere in Figure~\ref{fig:MNbS2_muonsite}] are identified at approximately (3/4, 2/3, 1/4) and other crystallographically equivalent positions [see sites D and E in Table~\ref{tab:sym_muonsite}]. 
A similar muon-stopping site has previously been suggested in~\crns~\cite{braam2015magnetic}.
The helical magnetic state produces near identical local field distributions for all the crystallographically equivalent positions in the undistorted structure, but slightly overestimate the precession frequency to be $110$~MHz.
However, after accounting for muon-induced distortions the local field distribution at site D results in a muon precession frequency approximately equal to $\nu_1$.
\begin{table}
\begin{tabular}{c|c|cc}
            &  & \multicolumn{2}{c}{Hyperfine Contact }~\\ 
       Material & Muon Site & \multicolumn{2}{c}{ Field Magnitude}~\\       
                 &               & (mT) & (MHz)\\
       \hline\hline
         \fens & A & $<2$ & $<0.3$\\
         & B & $<2$ & $<0.3$\\ 
            & C & $<2$ & $<0.3$\\
            & D & $510$ & $69$\\
            & E & $510$ & $69$\\
            \hline
         \vns & A & $<2$ & $<0.3$\\
         & B &  $<2$ & $<0.3$\\ 
            & C & $<2$ & $<0.3$\\
            & D & $1600$ & $220$\\
            & E & $1500$ & $200$\\
            \hline
         \mnns & A & - & -\\
         & B &  $<2$ & $<0.3$\\ 
            & C & $<2$ & $<0.3$\\
            & D & $300$ & $41$\\
            & E & $300$ & $41$\\
            \hline
        MnSi &  &$207$ & $28.0$\\
        \hline
        MnGe &  &$1080$ & $146$\\
\end{tabular}
\caption{Calculated hyperfine contact fields at the candidate muon sites for~\mnsall, along with experimentally reported coupling constants for a number of other materials~\cite{amato2014_MnSi,martin2016_MnGe}.
}
\label{tab:hyperfine}
\end{table}

Another possible explanation for this high frequency is the presence of a large hyperfine coupling at the muon site due to localised electron charge density~\cite{onuorah2018hf}.
This produces a hyperfine contact field [see Eq.~\ref{eq:hyperfine_contact}], which contributes to the local magnetic field at the muon site and the resulting muon precession frequency.
Using first-principles methods~\cite{walle1993_dft_hf} we estimate hyperfine coupling constants at the candidate muon sites in~\mnsall.
For sites B and C the contact fields in~\mnns~are similar to the other materials in the series [Table~\ref{tab:hyperfine}], and are too small to account for the size of local field responsible for $\nu_1$.
In contrast, the calculated contact field at sites D and E is significantly larger and, when accounted for in calculations of the local field distribution at the muon site, leads to precession frequencies that match $\nu_1$ for both magnetic structures.
Therefore, even when accounting for hyperfine contact fields, the higher-energy muon site must be realised in order to produce the observed precession frequency $\nu_1$.
Determining the occupation of muon stopping sites in materials is still an open question, so the realisation of unexpected sites (such as the high energy sites in~\mnns) is particular interesting.
We therefore propose that the~\mns~series provides an ideal test case for understanding the factors driving muon site occupation, as a change in only intercalant species alters the realised muon sites.

For the low energy site, the calculated field distribution is predominantly along the $c$ axis for the helical state and approximately along $[0.36, 0.05, 0.93]$ for the FM state. This means the amplitude $A_{2_{ab}}$ should be larger than the amplitude $A_{2_c}$
In contrast, the high energy site has a field distribution mostly confined within the $a$-$b$ plane for the helical state and approximately along $[0.5, 0.85, 0.1]$ for the FM state, so we expect that $A_{1_c}>A_{1_{ab}}$. 
Both of these amplitude relations are observed in our~\musr~measurements.
The relaxation seen due to slow magnetic dynamics, $A_{3_{ab}}$, can be primarily associated to the relaxation of muons in the high energy site, as the field is orientated in the $a$-$b$ plane for that site.

We conclude that, in contrast to other materials in the~\mns~series, the~\musr~spectra of~\mnns~consists of two proportional precession frequencies, where the high-frequency oscillation corresponds to the realisation of a higher-energy muon site. 
DFT calculations show that the magnetism of the ~\mns~series can be explained by electrons from the intercalated species filling  rigid-bands in the electronic structure~\cite{hawkhead2023intTMDC}. 
This leads to changes in the Fermi surfaces across series, such as an increase in Fermi surface nesting in the Fe and Mn materials compared to others in the series.
Features in the electronic structure, such as the pseudogap in~\crns~\cite{ghimire2013magnetic,hicken2022energy}, that are key in driving the magnetism of these materials are also strongly dependent on intercalant species, although the pseudogap, specifically, is not present in the $N=$~V, Fe and Mn materials discussed here.
However, electronic-structure considerations do suggest significant differences in the electronic energy landscape across the series, which might lead to the realisation of an additional muon-stopping site in~\mnns~that does not occur in other materials.

\section{Conclusion}
\label{Conclusion_sec}
The magnetic properties of~\mnsall~have been investigated making use of~\musr, an experimental technique sensitive to the local magnetic ordering and dynamics in materials, and complementary computational techniques.
In~\fens~muons are sensitive to the presence of both stripe and zig-zag magnetic phases, the changes in the population of which can be associated to components in our wTF measurements.
ZF~\musr~identifies transitions from a mixed phase to a stripe-dominated phase at $T_{c2}=40.2$~K and to paramagnetism at $T_{c1}=45$~K, where the observed oscillations can be attributed to muons implanting in regions of zig-zag ordering.
Moreover, the anisotropy in magnetic response seen in the longitudinal and transverse directions agrees with the orientation of the local field calculated at the candidate muon site determined in~\fens.
\vns~undergoes a transition to LRO at $T_c=52.7$~K, demonstrating a broad single frequency response which is compatible with both the proposed simple AFM and double-Q magnetic orderings.
Additionally, the low temperature transition at $\approx10$~K is seen as a peak in the dynamical fluctuation rate on the muon (MHz) timescale.
For~\mnns~a transition to LRO is seen at $T_c=44.2$~K, below which a ferromagnetic-like systems forms with ZF~\musr~revealing two oscillation frequencies that vary in fixed proportion.
The smaller frequency can be associated to muons implanting in the low energy candidate muon site seen across the series, but the high frequency response requires the realisation of a higher-energy muon site only seen in this material. 
This is supported by the orientation of the calculated local fields at both of the muon sites, and remains true even if the hyperfine interaction is accounted for.

We conclude that each of these materials demonstrates notably different magnetic behaviour.
Variations in magnetism are expected based on overall trends suggested by a rigid-band-filling model (which agrees with DFT calculations), but the formation of helical magnetic states, topological excitations and phase coexistence suggest more complex variations in the magnetic interactions.
Our results therefore indicate the broad range of magnetic phenomena produced using intercalation, highlighting the importance of the TMDCs as basis for tunable magnetic systems, and reaffirming~\musr~as a powerful technique for studying complex magnetism in materials.

\section{Acknowledgements}
Part of this work was carried out at the ISIS Neutron and Muon source, Rutherford Laboratory, UK, and S$\mu$S, Paul Scherrer Institute, Switzerland and we are grateful for the provision of beamtime. We also acknowledge travel support and (for A.H.-M) studentship support  from STFC-ISIS. 
This work is supported by EPSRC (UK) under grant EP/Z534067/1 and
also via studentship support for T.L.B and A.H.-M.
N.~P.~B acknowledges the support of the Durham Doctoral Scholarship. 
We thank Visagan Ravindran for discussions regarding the DFT calculations and acknowledge the support of Durham Hamilton HPC.
Research data will be made available via \textcolor{red}{XXX}, and the ISIS data can be found at https://doi.org/10.5286/ISIS.E.RB2220176.

\bibliography{bib.bib}

\end{document}